\documentclass[journal]{IEEEtran}

\ifCLASSINFOpdf
\else
   \usepackage[dvips]{graphicx}
\fi
\usepackage{url}

\usepackage{cite}
\usepackage{graphicx}
\usepackage{amsmath}
\usepackage{amssymb}
\usepackage{cleveref}
\usepackage{flushend}
\usepackage{multirow}
\usepackage{threeparttable}
\usepackage{booktabs}
\usepackage{adjustbox}
\usepackage{tablefootnote}
\usepackage{CJKutf8}
\crefname{equation}{Eq.}{Eqs.}
\usepackage[T1]{fontenc}
\begin{document}

\title{DAMOS: Learning Distortion-Aware Speech Quality Assessment through Explicit Distortion Localization
}

\author{
        Naiyuan~Li,
        Li~Dong,~\IEEEmembership{Member,~IEEE,}
        Diqun~Yan,~\IEEEmembership{Member,~IEEE,}

        \thanks{This work was supported by the National Natural Science Foundation of China (Grant Nos.~62571283, 62171244). (Corresponding author: Diqun Yan.)}
        \thanks{Naiyuan Li and Li Dong are with the College of Information Science and Engineering, Ningbo University, Ningbo, China (e-mail: 2411100281@nbu.edu.cn; dongli@nbu.edu.cn).}
        
         \thanks{Diqun Yan is with the College of Artificial Intelligence, Ningbo University of Finance and Economics, Zhejiang Provincial Collaborative Innovation Center for Digital Supply Chain and Artificial Intelligence of Bulk Commodities, and also with the College of Information Science and Engineering, Ningbo University, Ningbo, China (e-mail: yandiqun@nbufe.edu.cn; yandiqun@nbu.edu.cn).}

}

\markboth{IEEE Transactions on Audio, Speech, and Language Processing, Vol. x, No. x, x}
{Shell \MakeLowercase{\textit{et al.}}: Bare Demo of IEEEtran.cls for IEEE Journals}
\maketitle

\begin{abstract}

Automatic speech quality assessment aims to predict Mean Opinion Scores (MOS) consistent with human subjective perception and is essential for evaluating speech generation, enhancement, and communication systems. For speech signals, especially synthetic speech, distortions often occur locally, and overall perceptual quality is usually dominated by a small number of perceptually salient distortion regions. However, most existing methods are primarily optimized with utterance-level MOS, which provides only coarse-grained supervision and offer no explicit indication of where perceptually important distortions occur. To address this limitation, we introduce explicit distortion localization as auxiliary knowledge for speech quality assessment. We construct the first partially distorted speech dataset with frame-level distortion annotations and train a localization model to generate distortion cues. Building on these cues, we propose DAMOS, a distortion-aware speech quality assessment framework that integrates localization information into the MOS prediction pipeline. Experiments on multiple public benchmarks demonstrate that DAMOS consistently outperforms existing methods and exhibits strong cross-dataset generalization, validating the effectiveness of explicit distortion localization for speech quality assessment.

\end{abstract}

\begin{IEEEkeywords}
Speech quality assessment, mean opinion score, frame-level distortion, partially distorted speech dataset.
\end{IEEEkeywords}

\IEEEpeerreviewmaketitle

\section{Introduction}

\IEEEPARstart{S}{peech} quality assessment (SQA) has become increasingly important with the rapid development of speech generation, enhancement, and communication technologies. Reliable SQA not only affects user experience but also plays a critical role in evaluating and improving various speech-related systems \cite{4389058, 941023, bvcc}. For instance, SQA serves as an essential tool for assessing the performance of speech synthesis systems and measuring the effectiveness of speech enhancement algorithms \cite{10888446,mosnet, dnsmos}, enabling the development of high-quality and perceptually optimized speech technologies.

The mean opinion score (MOS) is the most widely used metric for SQA \cite{itu_p8001}, representing the overall perceptual quality of a speech utterance evaluated by human listeners. In conventional SQA datasets \cite{tcdvoip,tencent,pstn,somos,nisqa}, each speech sample is assigned a single MOS value that reflects the listener's overall impression of the entire utterance. This utterance-level scoring mechanism has become the primary supervision signal for existing SQA models. 

Although MOS provides an effective measurement of global speech quality and existing methods have achieved remarkable progress \cite{survey, voicemos2023, voicemos2024, audiomos2025}, 
such utterance-level supervision introduces a fundamental limitation for current SQA models. A single MOS value assigned to an entire utterance provides only coarse-grained guidance, but does not indicate which temporal regions contribute to the perceived degradation or where quality impairments occur, making it difficult for models to explicitly identify and emphasize perceptually important distortion regions. Models must instead infer the relationship between local distortions and overall quality from a single coarse label. This indirect supervision limits the model's ability to perceive local distortions. Moreover, human subjective evaluation of speech quality is rarely determined by the average contribution of all temporal segments; it is often dominated by a small number of perceptually salient distortion regions. Such a mismatch between utterance-level labels and local perceptual salience hampers the learning of quality representations aligned with human auditory perception.

This phenomenon is particularly pronounced in synthetic speech, where the instability of text-to-speech and voice conversion processes frequently introduces local distortions such as pronunciation errors, local speech collapse, vocoder artifacts, and transient anomalies. Although such distortions occupy only a short duration, they can disproportionately degrade overall listening quality, implying that the key factor determining speech quality is not merely how many distortions exist but where they occur and how they affect perceptual quality. We argue that what current SQA lacks is not a more complex network architecture but auxiliary knowledge that explicitly describes where distortions occur. Equipping models with such prior knowledge would allow them to focus on the key regions that truly affect perceptual quality. Introducing explicit distortion localization, however, raises two challenges: first, existing public SQA datasets provide only utterance-level MOS without frame-level distortion annotations; second, how to effectively integrate localization information into the MOS prediction pipeline---such that it enhances quality assessment rather than introducing noise---remains unexplored.

To address these challenges, we propose DAMOS, a distortion-aware framework based on explicit distortion localization. We first construct a partially distorted speech dataset with automatically generated frame-level annotations and train a localization model to provide fine-grained distortion cues. We then design a localization-guided quality prediction framework that integrates distortion cues with SSL features, enabling the model to focus on local regions that affect overall quality and learn more effective quality representations.

The main contributions of this paper are summarized as follows:

1) We introduce explicit distortion localization into speech quality assessment for the first time and use distortion location information as auxiliary knowledge to guide MOS prediction, providing a new research direction for fine-grained speech quality modeling.

2) The first partially distorted speech dataset with explicit distortion location annotations is constructed. This dataset provides supervision information for training distortion localization models and offers a new data foundation for future research on fine-grained speech quality modeling.

3) We propose DAMOS, a distortion-aware speech quality assessment framework that exploits explicit distortion localization information to guide the model in learning distortion-sensitive quality representations through progressive localization-aware modeling, thereby improving MOS prediction performance.

4) Extensive experiments validate the effectiveness of the proposed method. Results on multiple public benchmarks demonstrate that DAMOS consistently outperforms existing advanced methods and exhibits strong cross-dataset generalization ability.

\section{Explicit Distortion Localization}
\label{sec:detection}

The objective of explicit distortion localization is not to improve distortion detection itself, but to provide transferable distortion-aware supervision for downstream speech quality assessment. As discussed above, existing SQA datasets lack frame-level distortion annotations, making explicit localization information unavailable during training. To bridge this gap, we first construct a partially distorted speech dataset with automatically generated frame-level labels and then train a dedicated distortion localization model. The pretrained model subsequently provides frame-level distortion cues as auxiliary knowledge for the proposed distortion-aware MOS prediction framework.

\subsection{Partial-Distortion Dataset Construction}

Learning distortion localization requires frame-level supervision. Manually annotating distortion boundaries, however, is prohibitively expensive and highly subjective due to the ambiguous transition between clean and distorted speech. Instead of relying on manual annotation, we construct a partially distorted speech dataset with automatically generated frame-level labels.

The proposed dataset is designed according to two principles. First, distortion boundaries should be explicitly controllable, enabling accurate frame-level annotation without human intervention. Second, the synthesized speech should preserve realistic distortion characteristics so that the learned localization knowledge can generalize to downstream speech quality assessment.

To satisfy the first principle, clean speech utterances are selected from LibriSpeech \cite{librispeech}. All utterances are resampled to 16~kHz and randomly cropped to durations between 5 and 15 seconds. Utterances shorter than the target duration are cyclically repeated before cropping, so that the short utterances are preserved rather than discarded.

Before distortion injection, an energy-based voice activity detection (VAD) is applied using a 16~ms frame length and an 8~ms frame shift. Frames whose energy is more than 40~dB below the peak energy are removed. The detected speech regions are further contracted by 20~ms at both boundaries, and only segments longer than 0.3~s are retained. This preprocessing suppresses ambiguous transition regions around speech onsets and offsets, thereby improving the reliability of the generated frame-level annotations.

To satisfy the second principle, we synthesize diverse degradation patterns by introducing four categories of distortions according to their interaction with speech activity.

\textbf{Global distortions} modify the speech signal regardless of speech activity, including codec compression, codec tandeming, low-bitrate compression, resampling, and reverberation.

\textbf{Additive distortions} introduce environmental noise independently of speech activity, including additive white Gaussian noise (AWGN), pink noise, and real-world DNS noises \cite{dnsChallenge}. Prior to mixing, DNS noise recordings undergo the same energy validity check using a threshold of 35~dB below the peak energy to remove near-silent regions.

\textbf{Signal-dependent distortions} are only applied to VAD-active speech regions to ensure physical plausibility. This category includes modulation noise reference unit (MNRU), clipping, gain scaling, filtering, bandwidth compression, packet loss, phase randomization, time stretching, pitch shifting, spectral smoothing, Mel-spectrogram reconstruction, and vocoder mismatch.

To better approximate practical speech degradations, we further construct a \emph{combined distortion} subset by superimposing one additive distortion with one global or signal-dependent distortion. Approximately 60\% of these samples contain commonly observed degradation pairs, including DNS noise with reverberation, DNS noise with packet loss, and AWGN with clipping, while the remaining samples are randomly generated. In addition, a \emph{full-distortion} subset is constructed by applying one randomly selected distortion over the entire active speech region.

For each utterance, one to three non-overlapping distortion segments are randomly inserted. Each segment has a duration between 0.5~s and 3~s, with a minimum interval of 0.3~s between adjacent segments. The locations of distortion segments follow category-specific constraints to ensure both accurate boundary control and realistic degradation patterns.

Since the distortion locations are explicitly known during synthesis, frame-level binary labels can be automatically generated. Each utterance is divided into non-overlapping 160~ms frames. A frame is labeled as distorted if more than half of its duration overlaps with any injected distortion segment; otherwise, it is regarded as clean. This criterion improves robustness against slight temporal boundary mismatch between synthesized distortions and localization predictions.

Following the above procedure, we generate a dataset containing 15,000 utterances with an 8:1:1 train/validation/test split. Approximately 3\% of the samples contain clean speech without any injected distortion. Instead of adopting a uniform sampling strategy, the occurrence frequency of each distortion type is intentionally designed according to practical degradation priors. As summarized in Table~\ref{tab:dataset}, environmental noise, reverberation, packet loss, and combined distortions are assigned relatively higher proportions, while less common degradations are sampled with lower probabilities. Such a non-uniform distribution encourages the localization model to learn diverse distortion characteristics and improves its transferability to downstream speech quality assessment.
\begin{table}[t]
\centering
\caption{Distortion type distribution in the synthetic dataset (actual generated statistics).}
\label{tab:dataset}
\resizebox{\columnwidth}{!}{%
\begin{tabular}{@{}llcc@{}}
\toprule
Class & Type & Samples & Ratio (\%) \\
\midrule
--  & Clean                       &   406 &  2.7 \\
\midrule
A1  & Codec (single)              & 1,327 &  8.8 \\
A2  & Codec tandem                &   693 &  4.6 \\
A3  & Low-bitrate compression     &   871 &  5.8 \\
A4  & Resampling                  &   565 &  3.8 \\
A5  & Reverberation               & 1,411 &  9.4 \\
\midrule
B1  & AWGN                        &   584 &  3.9 \\
B2  & Pink noise                  &   281 &  1.9 \\
B3  & Real-world noise (DNS)      & 1,620 & 10.8 \\
\midrule
C1  & MNRU                        &   591 &  3.9 \\
C2  & Clipping                    &   839 &  5.6 \\
C3  & Gain scaling                &   274 &  1.8 \\
C4  & Filtering                   &   736 &  4.9 \\
C5  & Bandwidth compression       &   579 &  3.9 \\
C6  & Packet loss                 & 1,096 &  7.3 \\
C7  & Phase randomization         &   288 &  1.9 \\
C8  & Time stretching             &   278 &  1.9 \\
C9  & Pitch shifting              &   312 &  2.1 \\
C10 & Spectral smoothing          &   256 &  1.7 \\
C11 & Mel-spectrogram reconstruction &   284 &  1.9 \\
C12 & Vocoder mismatch            &   293 &  2.0 \\
\midrule
E   & Combined distortion         & 1,416 &  9.4 \\
\midrule
\multicolumn{2}{l}{\textbf{Total}} & \textbf{15,000} & \textbf{100.0} \\
\bottomrule
\end{tabular}%
}
\end{table}

\subsection{Distortion Localization Model}

The objective of the distortion localization model is to learn transferable distortion-aware representations rather than directly predict speech quality. Unlike utterance-level MOS prediction, distortion localization focuses on identifying temporal boundaries between clean and degraded speech regions.

Considering its strong capability in modeling boundary transitions, we adopt the Boundary-Aware Model (BAM) \cite{bam}, originally developed for partial spoofing detection, as the localization backbone. All hyperparameters follow the original implementation.

Given an input waveform $\mathbf{x}$, BAM predicts a frame-level binary distortion localization map:

\begin{equation}
\mathbf{M}
=
f_{\mathrm{BAM}}(\mathbf{x})
=
[m_1,m_2,\ldots,m_T]
\in
\{0,1\}^{T},
\label{eq:Mask}
\end{equation}
where $f_{\mathrm{BAM}}(\cdot)$ denotes the BAM model. Each binary value $m_t=1$ indicates that a distortion is detected in frame $t$, while $m_t=0$ represents a clean frame.

The localization model is trained independently on the proposed partially distorted dataset. During speech quality assessment, all BAM parameters remain frozen and are never optimized using MOS supervision. The predicted localization map is only used as explicit distortion-aware guidance, which is aligned with the temporal resolution of SSL representations and incorporated into the proposed distortion-aware MOS prediction framework. This design allows distortion localization and quality prediction to focus on complementary objectives.

\begin{figure*}[!t]
  \centering
  \includegraphics[width=\textwidth]{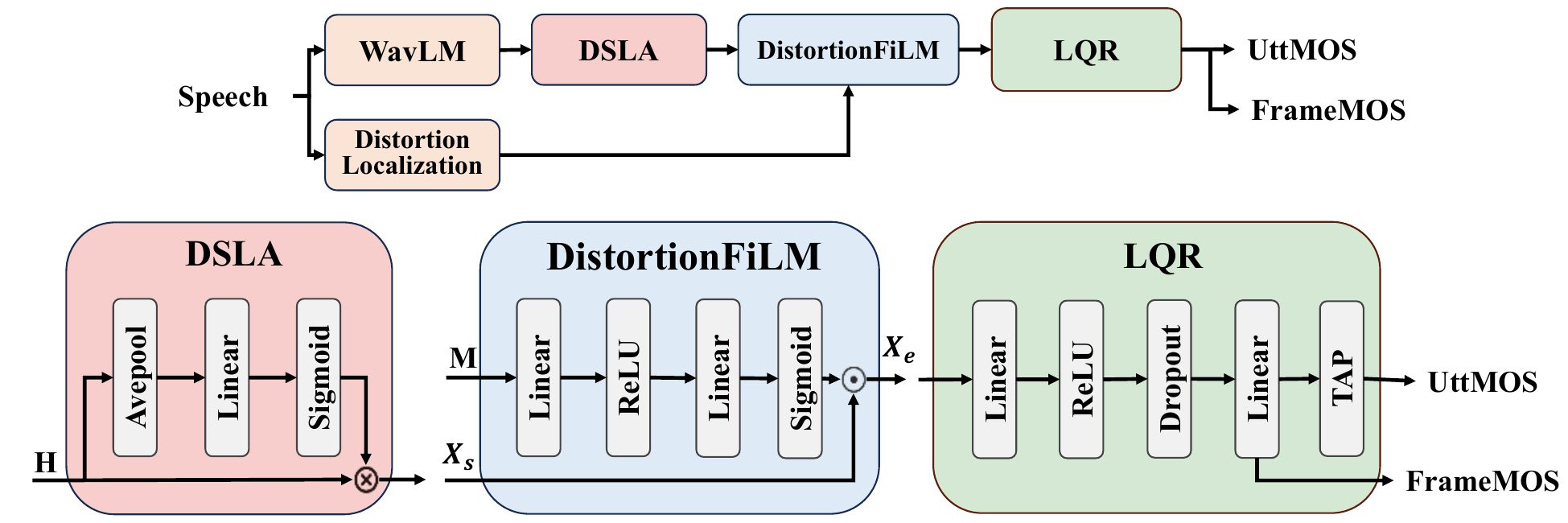}
  \caption{Architecture of the proposed DAMOS.}
  \label{fig:architecture}
\end{figure*}

\section{Distortion-Aware MOS Prediction}
\label{sec:method}

\subsection{Overview}

Section~\ref{sec:detection} introduces an explicit distortion localization model that provides reliable frame-level localization cues. The remaining challenge is how to effectively exploit such localization knowledge for utterance-level MOS prediction. A straightforward solution is to directly fuse the predicted localization map with the quality representation. However, such one-stage fusion introduces distortion information at only a single point of the network and cannot fully exploit its guidance throughout the quality prediction process.

We argue that explicit distortion localization should participate in the entire MOS prediction pipeline rather than serving as an auxiliary feature at the fusion stage alone. Specifically, localization knowledge should first guide the extraction of distortion-sensitive representations, then dynamically condition feature learning to emphasize perceptually degraded regions, and finally be preserved until quality estimation so that localized distortion cues can contribute directly to utterance-level MOS prediction. Following this principle, we propose a distortion-aware MOS prediction framework consisting of three successive stages, namely Distortion-Sensitive Layer Adaptation (DSLA), Distortion-Guided Feature Modulation (DistortionFiLM), and Localized Quality Regression (LQR). As illustrated in Fig.~\ref{fig:architecture}, these three components progressively discover, inject, and preserve explicit distortion localization knowledge throughout the entire quality prediction pipeline.

\subsection{Distortion-Sensitive Layer Adaptation}

The first step of the proposed framework is to discover representations that are more sensitive to speech distortions. Recent studies have shown that hidden representations of self-supervised speech models encode hierarchical information across transformer layers \cite{layer1, layer2, layer3}, ranging from low-level acoustic characteristics to high-level semantic content. Consequently, speech-quality-related information is not uniformly distributed throughout the network but is instead scattered across multiple hidden layers.

However, most existing SSL-based speech quality assessment methods simply adopt the final-layer representation or aggregate multiple layers using fixed strategies such as averaging or concatenation \cite{sslmos, utmos, mosanet}.
Such approaches implicitly assume that all hidden layers contribute equally to quality assessment, potentially overlooking intermediate representations that are more responsive to speech distortions. Moreover, different SQA datasets exhibit distinct acoustic characteristics and distortion distributions, meaning that the most informative transformer layer may vary across evaluation scenarios. Consequently, fixed layer selection or static aggregation may become suboptimal on unseen data, limiting model robustness and cross-dataset generalization. Since the subsequent feature modulation explicitly exploits distortion localization knowledge, adaptively identifying distortion-sensitive representations becomes a prerequisite for effective distortion-aware feature learning.

To this end, we propose a lightweight Distortion-Sensitive Layer Adaptation (DSLA) module that automatically estimates the importance of each transformer layer and adaptively aggregates distortion-sensitive representations. The hidden representations $\mathbf{H}$ are extracted from all $L$ transformer layers:

\begin{equation}
\mathbf{H}
=
[\mathbf{h}_1,\ldots,\mathbf{h}_L]
\in
\mathbb{R}^{L\times T\times D},
\end{equation}
where $T$ and $D$ denote the temporal length and feature dimension respectively. 
Temporal average pooling $\operatorname{AvgPool}_{T}$ is first applied over the temporal axis $T$. The pooled feature is subsequently passed through a linear layer $\operatorname{FC}$, followed by a sigmoid activation $\sigma(\cdot)$ to produce
normalized layer-importance weights $\mathbf{w}$:

\begin{equation}
\mathbf{w}
=
\sigma
\left(
\mathrm{FC}
\left(
\mathrm{AvgPool}_{T}
(\mathbf{H})
\right)
\right),
\label{eq:dsla_weight}
\end{equation}
where $\mathbf{w} = [w_1, w_2, \ldots, w_L] \in \mathbb{R}^{L}$ 
denotes the learned importance weights assigned to individual transformer layers, and $w_l$ is the importance of layer $l$.

Finally, the distortion-sensitive representation is obtained as a weighted aggregation of all hidden representations:

\begin{equation}
\mathbf{x}_s
=
\sum_{l=1}^{L}
w_l
\mathbf{h}_l.
\label{eq:dsla}
\end{equation}

Unlike fixed layer aggregation strategies, DSLA automatically discovers transformer layers that are more informative for distortion perception and adaptively emphasizes their contributions. The resulting representation serves as the foundation for the subsequent distortion-guided feature conditioning stage.

\subsection{Distortion-Guided Feature Modulation}

After identifying distortion-sensitive representations, the remaining challenge is how to effectively inject explicit distortion localization knowledge into the quality prediction process. Without explicit localization cues, the model must implicitly infer distortion locations from a single utterance-level quality score. However, speech distortions are typically confined to only a small portion of an utterance, whereas Transformer-based SSL encoders perform long-range contextual modeling over the entire speech sequence. Consequently, localized distortion cues become progressively coupled with abundant perceptually clean regions, making degraded frames increasingly difficult to distinguish---a phenomenon we refer to as local-global coupling \cite{frame1, frame2}. This coupling limits the model's ability to exploit localized distortion information under utterance-level supervision alone.

To explicitly alleviate this limitation, we propose DistortionFiLM, a lightweight feature modulation module that injects distortion localization knowledge into the quality representation. Unlike conventional feature fusion, DistortionFiLM conditions feature learning on the predicted localization map, allowing explicit localization cues to dynamically modulate feature responses without increasing the representation dimensionality.

Specifically, the pretrained localization model predicts a binary localization map indicating whether each temporal frame is distorted. Since the temporal resolution of the localization model differs from that of the SSL encoder, the predicted localization map is first aligned to the SSL feature sequence using nearest-neighbor interpolation. The aligned localization map is subsequently projected into the feature space through a lightweight multilayer perceptron (MLP) followed by a sigmoid activation $\sigma(\cdot)$ to generate a frame-wise modulation gate $\mathbf{g}$:

\begin{equation}
\mathbf{g}
=
\sigma
\left(\mathbf{W_2}
\mathrm{ReLU}
(\mathbf{W_1}\mathbf{M})
\right),
\label{eq:film}
\end{equation}
where $\mathbf{W}_1 \in \mathbb{R}^{H \times 1}$ and $\mathbf{W}_2 \in \mathbb{R}^{H \times D}$ are the weights of the MLP. The mask
$\mathbf{M}$ is obtained from Eq.~\eqref{eq:Mask}.

The distortion-sensitive representation $\mathbf{x}_e$is then conditioned by:

\begin{equation}
\mathbf{x}_e
=
\mathbf{g}
\odot
\mathbf{x}_s,
\label{eq:modulation}
\end{equation}
where $\odot$ denotes element-wise multiplication.

Through this distortion-guided conditioning process, feature responses corresponding to distorted regions are selectively emphasized, while those from distortion-irrelevant regions are relatively suppressed. Consequently, explicit localization knowledge is injected into the quality representation before MOS regression, producing distortion-aware features that are more discriminative for speech quality assessment.

\subsection{Localized Quality Regression}

After distortion-aware feature conditioning, the enhanced local representations should be preserved until the final quality estimation. However, several works first aggregate frame-level representations through temporal average pooling and subsequently perform utterance-level regression \cite{sslmos, ramp, meta, t05}. Such early temporal aggregation inevitably mixes distortion-enhanced frames with abundant distortion-irrelevant regions, weakening the contribution of localized degradation cues.

To preserve explicit distortion localization knowledge throughout the prediction process, we perform quality regression before temporal aggregation. Specifically, the distortion-aware representation is denoted as $\mathbf{x}_e \in \mathbb{R}^{T \times D}$.

Then it is fed into a lightweight prediction head consisting of two linear layers with ReLU activation and dropout ($p=0.1$). The prediction head independently estimates the quality of every temporal frame.
The frame-level quality scores $\mathrm{FrameMOS_t}$are first estimated by the prediction head:

\begin{equation}
\mathrm{FrameMOS_t}
=
\mathbf{W}_4
\mathrm{Dropout}
\left(
\mathrm{ReLU}
(\mathbf{W}_3\mathbf{x}_{e,t})
\right),
\label{eq:frame_mos}
\end{equation}
where $\mathrm{FrameMOS_t}$ denotes the predicted MOS score of frame $t$. 
$\mathbf{W}_3 \in \mathbb{R}^{H \times D}$ and 
$\mathbf{W}_4 \in \mathbb{R}^{1 \times H}$ are the weights of the first and second linear layers in the quality prediction head, respectively. And the $\mathbf{x}_{e,t}$ denotes the distortion-aware representation of the frame $t$.

The utterance-level MOS is then obtained through temporal average pooling (TAP):

\begin{equation}
{\mathrm{UttMOS}}
=
\frac{1}{T}
\sum_{t=1}^{T}
\mathrm{FrameMOS_t}.
\label{eq:tap}
\end{equation}

The entire framework is optimized using the mean squared error (MSE) between the predicted and ground-truth utterance-level MOS. Although only utterance-level supervision is available during training, delaying temporal aggregation until after frame-level regression allows localized distortion cues to directly contribute to quality estimation. Consequently, explicit distortion localization knowledge introduced by DistortionFiLM is preserved throughout the prediction pipeline rather than being diluted during early feature aggregation. The distortion-aware representation obtained here is preserved until the final utterance-level estimation, completing the progressive discover--inject--preserve pipeline.

\section{Experiments}
\label{sec:exp}
\subsection{Experimental Setup}
\subsubsection{Dataset}
We evaluate the proposed method on the BVCC dataset \cite{bvcc}, which contains synthetic speech samples from Blizzard Challenge \cite{blizzard} and Voice Conversion Challenge \cite{vcc2016,vcc2018,vcc2020} with corresponding MOS scores. All waveforms are resampled to 16\,kHz. We follow the predefined train/validation/test split of 4974/1066/1066 for BVCC. We use standard performance evaluation metrics, mean-
squared-error (MSE), linear-correlation-coefficient (LCC),and Spearman’s rank-correlation-coefficient (SRCC) to assess our models at the utterance and system levels. During training, utterances are randomly cropped to a maximum length of 8 seconds; during evaluation, center cropping is applied with zero-padding for shorter utterances.

\subsubsection{Implementation Details}
We adopt WavLM-Large \cite{wavlm} as the default SSL backbone. The model is optimized using Adam with grouped learning rates: SSL parameters use $lr=1\times10^{-6}$ while other parameters use $lr=1\times10^{-4}$. Weight decay is set to $1\times10^{-4}$. We apply a StepLR scheduler with step size 10 and decay factor 0.1. The model is trained for 30 epochs with a batch size of 16 on a single GPU. We select the checkpoint with the highest validation SRCC for testing.

\subsection{Results}
\subsubsection{Comparison with State-of-the-Art Methods}

To evaluate the effectiveness of the proposed DAMOS, we compare it with representative state-of-the-art SQA methods. Table~\ref{tab:main} reports the comparison results on the BVCC dataset at both utterance and system levels. In addition, we also evaluate it on several non-synthetic speech datasets. Evaluations on non-synthetic speech benchmarks, including PSTN \cite{pstn}, Tencent \cite{tencent}, and NISQA \cite{nisqa}, are presented in Table~\ref{tab:multi_dataset}. All methods are trained and evaluated in an in-domain setting on each respective dataset.

Except for SLL-Conformer, all baseline methods are reproduced using their publicly available implementations under our experimental environment. To ensure a fair comparison, all reproduced models are trained using the same dataset partition strategy as DAMOS.

As shown in Table~\ref{tab:main}, at the utterance level, compared with UTMOS, which represents a strong SSL-based baseline, DAMOS improves SRCC from 0.878 to 0.885 while simultaneously reducing prediction errors. Similar improvements are also observed compared with SSl-MOS and other quality assessment approaches, demonstrating that the proposed distortion-aware modeling strategy can further enhance modern SQA systems.

At the system level, although the performance gap between different methods becomes smaller due to the reduced evaluation variance, DAMOS still achieves the most accurate system-level quality ranking and the lowest prediction error, indicating that the proposed framework effectively captures quality differences among different speech generation systems.

As shown in Table~\ref{tab:multi_dataset}, DAMOS achieves consistent improvements across diverse speech quality assessment scenarios, including telephone speech, conversational speech, and real-world noisy conditions. In particular, DAMOS obtains the best performance on most datasets and achieves superior average performance compared with existing approaches. These results demonstrate that the proposed framework is not limited to a specific benchmark but can be effectively applied to different distortion characteristics and acoustic environments.

To further evaluate the generalization ability of the proposed method across various datasets, we compare DAMOS with representative speech quality assessment methods on seven additional datasets covering multiple application scenarios, including speech enhancement, speech communication, synthetic speech, and real recording. All models are trained on the BVCC dataset. The results are summarized in Table~\ref{tab:srcc_comparison}. DAMOS achieves the highest SRCC values on the TCD-VoIP, NISQA\_FOR, and SOMOS datasets and the best average performance across all evaluation datasets. On several other datasets, DAMOS achieved the second-best performance. These results demonstrate that the proposed distortion-aware modeling strategy generalizes well to different distortion types and recording conditions, rather than being optimized specifically for a single benchmark.

\begin{table}[!t]
  \centering
  \caption{Performance comparison on the BVCC test set.}
  \label{tab:main}
  \begin{threeparttable}
  \begin{adjustbox}{width=0.48\textwidth}
  \begin{tabular}{lcccccc}
    \toprule
    & \multicolumn{3}{c}{\textbf{Utterance level}} & \multicolumn{3}{c}{\textbf{System level}} \\
    \cmidrule(lr){2-4} \cmidrule(lr){5-7}
    \textbf{Method} & \textbf{MSE} & \textbf{SRCC} & \textbf{LCC} & \textbf{MSE} & \textbf{SRCC} & \textbf{LCC} \\
    \midrule
    
    LDNet \cite{ldnet} & 0.332 & 0.784 & 0.788 & 0.173 & 0.870 & 0.876 \\
    NISQA \cite{nisqa} & 0.380 & 0.784 & 0.788 & 0.201 & 0.867 & 0.871 \\
    SSL-MOS \cite{sslmos} & 0.252 & 0.870 & 0.869 &  0.200 & 0.928 & 0.924 \\
    UTMOS \cite{utmos} & 0.408 & 0.878 & 0.878 & 0.247 & 0.930 & 0.929 \\
    MOSA-Net+ \cite{mosanet} & 0.324 & 0.800 & 0.801 & 0.188 & 0.885 & 0.882 \\
    DNSMOS Pro \cite{dnsmospro} & 0.345 & 0.780 & 0.784 & 0.155 & 0.874 & 0.880 \\
    DeepMOS \cite{deepmos} & 0.584 & 0.733 & 0.739 & 0.423 & 0.850 & 0.852 \\
    
    \midrule
    \textbf{Proposed} & \textbf{0.191} & \textbf{0.885} & \textbf{0.884} & \textbf{0.100} & \textbf{0.931} & \textbf{0.931} \\
    \bottomrule
  \end{tabular}
  \end{adjustbox}
  \end{threeparttable}
\end{table}

\begin{table*}[t]
\centering
\caption{performance comparison on the NISQA, PSTN and Tencent datasets.}
\label{tab:multi_dataset}
\begin{threeparttable}
\renewcommand{\TPTminimum}{\textwidth}
\resizebox{\textwidth}{!}{

\begin{tabular}{lcccccccccc}
\toprule
\multirow{2}{*}{Method}
& \multicolumn{2}{c}{PSTN}
& \multicolumn{2}{c}{Tencent}
& \multicolumn{2}{c}{NISQA\_FOR}
& \multicolumn{2}{c}{NISQA\_LIVETALK}
& \multicolumn{2}{c}{NISQA\_P501} \\
& SRCC & LCC
& SRCC & LCC
& SRCC & LCC
& SRCC & LCC
& SRCC & LCC \\
\midrule
NISQA
& 0.819 & 0.818
& 0.943 & 0.946
& 0.876 & 0.891
& 0.755 & 0.778
& 0.907 & 0.904 \\

SSL-MOS
& 0.816 & 0.812
& 0.959 & 0.960
& 0.859 & 0.857
& 0.674 & 0.696
& 0.902 & 0.895 \\

MOSA-Net+
& 0.816 & 0.812
& 0.936 & 0.937
& 0.863 & 0.863
& 0.633 & 0.633
& 0.876 & 0.887 \\

DNSMOS Pro
& 0.799 & 0.797
& 0.918 & 0.923
& 0.754 & 0.760
& 0.519 & 0.512
& 0.864 & 0.824 \\

SLL-Conformer \cite{SLLConformer}
& 0.825 & 0.822
& 0.952 & 0.954
& 0.904 & 0.914
& 0.759 & 0.778
& \textbf{0.926} & \textbf{0.928} \\

DAMOS (Ours)
& \textbf{0.828} & \textbf{0.829}
& \textbf{0.971} & \textbf{0.972}
& \textbf{0.924} & \textbf{0.924}
& \textbf{0.868} & \textbf{0.881}
& 0.919 & 0.914 \\
\bottomrule
\end{tabular}}
\begin{tablenotes}
\footnotesize

\item[1] The results of SLL-Conformer are directly obtained from its original paper, where the authors
adopted the same dataset split protocol as ours;

\item[2] SLL-Conformer utilizes additional training data
from the NISQA, PSTN, and Tencent datasets during training, which may provide an advantage in terms of data diversity and generalization capability;

\item[3] LDNet and UTMOS
require listener identity information (listener-id) during training, which is only available in the BVCC dataset. Therefore,
these two models are reproduced and evaluated only on BVCC;

\end{tablenotes}

\end{threeparttable}
\end{table*}

\begin{table*}[t]
  \centering
  \caption{Cross-corpus generalization results: SRCC comparison across datasets.}
  \label{tab:srcc_comparison}
  \small
  \setlength{\tabcolsep}{4pt}
  \begin{tabular}{lcccccccc}
    \toprule
    \textbf{Model} & \textbf{TCD-VoIP} \cite{tcdvoip} & \textbf{NISQA\_P501} & \textbf{NISQA\_LIVETALK} & \textbf{NISQA\_FOR} & \textbf{Tencent} & \textbf{PSTN} & \textbf{SOMOS} \cite{somos} & \textbf{Avg.} \\
    \midrule

    LDNet 
    & 0.734 & 0.611 & -0.092 & 0.393 & 0.328 & 0.578 & -0.030 & 0.360 \\

    NISQA 
    & 0.561 & 0.519 & 0.236 & 0.359 & 0.241 & 0.433 & 0.111 & 0.351 \\

    SSL-MOS 
    & 0.750 & 0.847 & 0.680 & 0.737 & 0.675 & 0.736 & \textbf{0.435} & 0.694 \\

    UTMOS 
    & 0.775 & \textbf{0.890} & \textbf{0.792} & \underline{0.798} & 0.714 & \textbf{0.749} & 0.307 & \underline{0.718} \\

    MOSA-Net+ 
    & \underline{0.828} & 0.816 & 0.629 & 0.685 & \textbf{0.755} & 0.736 & 0.387 & 0.691 \\

    DNSMOS Pro 
    & 0.554 & 0.584 & 0.147 & 0.275 & 0.465 & 0.576 & 0.055 & 0.379 \\

    DeepMOS 
    & 0.580 & 0.659 & -0.150 & 0.394 & 0.532 & 0.651 & 0.092 & 0.394 \\



    \textbf{DAMOS}
    & \textbf{0.860} & \underline{0.856} & \underline{0.781} & \textbf{0.808} & \underline{0.742} & \underline{0.738} & \textbf{0.435} & \textbf{0.746} \\
    \bottomrule
  \end{tabular}
\end{table*}

\subsubsection{Ablation Study}

To better understand the contribution of each proposed component, we conduct ablation studies on the BVCC dataset. As shown in Table~\ref{tab:ablation}, removing any component consistently degrades the overall performance, demonstrating that the three modules are complementary.

\textbf{Effect of DistortionFiLM.}
Removing DistortionFiLM leads to a noticeable degradation on both utterance-level and system-level metrics. In particular, the utterance-level SRCC decreases from \textbf{0.885} to \textbf{0.877}, while the MSE increases from \textbf{0.191} to \textbf{0.205}. This result indicates that explicitly injecting distortion localization cues effectively guides the model to emphasize perceptually degraded regions instead of relying solely on utterance-level supervision.

\textbf{Effect of DSLA.}
Replacing the proposed Distortion-Sensitive Layer Adaptation with the conventional final-layer representation causes the largest performance degradation among all ablations. The utterance-level SRCC drops to \textbf{0.853}, accompanied by a substantial increase in prediction error. This observation suggests that speech-quality-related information is distributed across multiple hidden layers of the SSL encoder rather than being concentrated in the final layer. Adaptive layer aggregation is therefore essential for extracting distortion-sensitive representations.

\textbf{Effect of LQR.}
When the proposed Localized Quality Regression is replaced by the conventional utterance-level regression strategy, performance also decreases consistently. This result demonstrates that performing frame-level quality regression before temporal aggregation better preserves localized distortion information introduced by DistortionFiLM. Consequently, distortion-aware representations can contribute more effectively to the final utterance-level MOS estimation.

Overall, these ablation results verify that the proposed framework benefits from all three components. DSLA first discovers distortion-sensitive representations, DistortionFiLM injects explicit localization knowledge into feature learning, and LQR preserves localized quality information until the final prediction. Their combination produces the best overall performance.

\begin{table}[!t]
  \centering
  \caption{Ablation study on the BVCC test set.}
  \label{tab:ablation}
  \begin{threeparttable}
  \begin{adjustbox}{width=0.48\textwidth}
  \begin{tabular}{lcccccc}
    \toprule
    & \multicolumn{3}{c}{\textbf{Utterance level}} & \multicolumn{3}{c}{\textbf{System level}} \\
    \cmidrule(lr){2-4} \cmidrule(lr){5-7}
    \textbf{Configuration} & \textbf{MSE} & \textbf{SRCC} & \textbf{LCC} & \textbf{MSE} & \textbf{SRCC} & \textbf{LCC} \\
    \midrule
    w/o DistortionFiLM & 0.205 & 0.877 & 0.877 & 0.118 & 0.924 & 0.925 \\
    w/o DSLA (last layer only) & 0.252 & 0.853 & 0.855 & 0.119 & 0.921 & 0.923 \\
    w/o LQR  & 0.206 & 0.876 & 0.875 & 0.112 & 0.925 & 0.926 \\
    \midrule
    \textbf{Full model} & \textbf{0.191} & \textbf{0.885} & \textbf{0.884} & \textbf{0.100} & \textbf{0.931} & \textbf{0.931} \\
    \bottomrule
  \end{tabular}
  \end{adjustbox}
  \end{threeparttable}
\end{table}

\begin{table*}[t]
\centering
\caption{Analysis of different hidden-layer aggregation strategies. All models evaluated using utterance-level SRCC.}
\label{tab:layer_analysis}
\resizebox{\textwidth}{!}{
\begin{tabular}{lccccccccc}
\toprule
Method
& BVCC
& TCD-VoIP
& NISQA\_P501
& NISQA\_LIVETALK
& NISQA\_FOR
& Tencent
& PSTN
& SOMOS
& Avg \\
\midrule

Last Hidden Layer
& 0.853
& 0.735
& 0.826
& 0.333
& 0.480
& 0.687
& 0.630
& 0.380
& 0.616
\\

Best Hidden Layer
& \textbf{0.887}
& \underline{0.851}
& 0.827
& 0.727
& 0.767
& 0.719
& \underline{0.740}
& \underline{0.405}
& 0.740
\\

Average of All Layers
& 0.876
& 0.815
& \underline{0.855}
& \underline{0.759}
& \textbf{0.825}
& \textbf{0.791}
& \textbf{0.742}
& 0.399
& 0.758
\\

\textbf{DAMOS (DSLA)}
& \underline{0.885}
& \textbf{0.860}
& \textbf{0.856}
& \textbf{0.781}
& \underline{0.808}
& \underline{0.742}
& 0.738
& \textbf{0.435}
& \textbf{0.763}
\\

\bottomrule
\end{tabular}}
\end{table*}

\begin{figure*}[t]
    \centering
    \includegraphics[width=\textwidth]{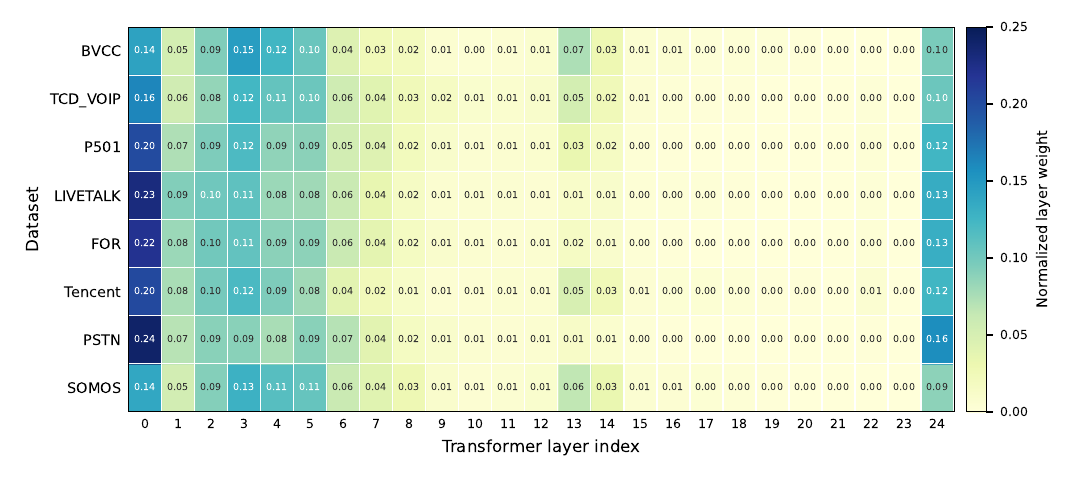}
    \caption{Average normalized layer weights learned by the proposed Distortion-Sensitive Layer Adaptation (DSLA) on different speech quality assessment datasets.}
    \label{fig:layer_weight}
\end{figure*}
\subsubsection{Analysis of Distortion-Sensitive Layer Adaptation}

The ablation study in Table~\ref{tab:ablation} demonstrates that removing DSLA causes the largest performance degradation among all proposed components, highlighting its importance in extracting effective quality representations. To further investigate the effectiveness of DSLA, we compare it with several commonly used hidden-layer selection and aggregation strategies, including the final hidden layer, the best-performing single hidden layer, and uniform averaging of all hidden layers.

Specifically, models using different layer aggregation strategies are trained on the BVCC training set and evaluated on the BVCC test set as well as seven additional speech quality assessment benchmarks. For the best hidden-layer strategy, we independently train models using each hidden layer representation and select the layer achieving the highest validation performance, which corresponds to the fifth hidden layer of the SSL encoder. The comparison results are presented in Table~\ref{tab:layer_analysis}.

The proposed DSLA achieves the best average performance across all evaluation datasets, demonstrating its superior ability to extract generalizable quality representations. Compared with using only the final hidden layer, selecting the fifth hidden layer provides significant improvements on the in-domain BVCC test set, indicating that intermediate SSL layers contain more informative representations for speech quality assessment. However, this fixed layer strategy shows degraded performance on several unseen datasets compared with both layer averaging and DSLA. This observation suggests that although a specific hidden layer may be optimal for a particular dataset, its representation may not generalize well under different distortion distributions and acoustic conditions.

Meanwhile, simple averaging of all hidden layers also fails to achieve optimal performance, despite exploiting information from the entire SSL encoder. This indicates that hidden layers contribute unequally to speech quality assessment, and that determining the appropriate contribution of each layer, rather than treating them uniformly, is essential for robust quality prediction. Notably, uniform averaging slightly outperforms DSLA on a few datasets (e.g., NISQA\_FOR, Tencent, PSTN), suggesting that adaptive weighting does not universally dominate simple averaging, though DSLA remains superior on average.

To further understand the behavior of DSLA, we analyze the learned layer importance weights on different speech quality assessment datasets, as illustrated in Fig.~\ref{fig:layer_weight}. The results reveal a consistent pattern across all datasets: the learned weights concentrate primarily on the early acoustic layers and the final layer of the SSL encoder, indicating that low-level acoustic characteristics and high-level contextual representations both provide generally important information for speech quality assessment. In contrast, the weighting of intermediate layers varies noticeably across datasets---for instance, BVCC and SOMOS assign substantially higher weights to the 13th and 14th layers, whereas PSTN does not exhibit a similar preference. This suggests that the relevance of intermediate representations depends more strongly on dataset-specific factors such as distortion type, recording condition, and speech characteristics, whereas the reliance on early and final layers reflects a more universal property of speech quality perception.

Consequently, fixed layer selection or uniform aggregation strategies may fail to consistently capture the most relevant representations across diverse evaluation scenarios. DSLA addresses this by dynamically estimating the contribution of each hidden layer according to the input, adaptively balancing the generally useful early and final layers with the dataset-dependent intermediate representations that fixed strategies cannot flexibly exploit. This adaptivity offers an intuitive explanation for the improved cross-dataset generalization observed in Table~\ref{tab:layer_analysis}.

\begin{table*}[htbp]
\centering
\caption{Results of Per-distortion-class frame-level and boundary detection metrics}
\label{tab:per_distortion_metrics}
\resizebox{\textwidth}{!}{%
\begin{tabular}{lcccccccccc}
\toprule
\multirow{2}{*}{Class} & \multicolumn{5}{c}{Frame-level} & \multicolumn{5}{c}{Boundary} \\
\cmidrule(lr){2-6} \cmidrule(lr){7-11}
 & EER(\%) & F1 & Acc & Prec & Rec & EER(\%) & F1 & Acc & Prec & Rec \\
\midrule
all         & 15.77 & 0.807 & 0.903 & 0.963 & 0.695 & 13.61 & 0.742 & 0.976 & 0.887 & 0.638 \\
\midrule
A1          & 16.40 & 0.819 & 0.906 & 0.976 & 0.705 & 12.74 & 0.762 & 0.979 & 0.909 & 0.656 \\
A2          & 14.57 & 0.775 & 0.894 & 0.965 & 0.648 & 16.35 & 0.683 & 0.972 & 0.835 & 0.578 \\
A3          & 12.87 & 0.838 & 0.920 & 0.944 & 0.753 & 12.79 & 0.802 & 0.981 & 0.934 & 0.703 \\
A4          & 19.27 & 0.767 & 0.897 & 0.953 & 0.642 & 9.47  & 0.793 & 0.983 & 0.873 & 0.727 \\
A5          & 17.23 & 0.798 & 0.901 & 0.939 & 0.694 & 16.89 & 0.681 & 0.970 & 0.866 & 0.561 \\
B1          & 15.41 & 0.817 & 0.909 & 0.975 & 0.703 & 15.10 & 0.757 & 0.976 & 0.944 & 0.632 \\
B2          & 9.80  & 0.857 & 0.925 & 0.981 & 0.760 & 11.67 & 0.694 & 0.967 & 0.797 & 0.615 \\
B3          & 17.77 & 0.779 & 0.890 & 0.961 & 0.655 & 13.76 & 0.719 & 0.974 & 0.902 & 0.598 \\
C1          & 11.97 & 0.867 & 0.925 & 0.944 & 0.802 & 11.41 & 0.772 & 0.975 & 0.865 & 0.697 \\
C10         & 10.41 & 0.791 & 0.900 & 0.963 & 0.671 & 13.27 & 0.743 & 0.974 & 0.899 & 0.633 \\
C11         & 21.22 & 0.727 & 0.868 & 0.966 & 0.583 & 17.17 & 0.699 & 0.973 & 0.891 & 0.576 \\
C12         & 12.70 & 0.869 & 0.940 & 0.974 & 0.785 & 12.05 & 0.802 & 0.982 & 0.949 & 0.694 \\
C2          & 13.68 & 0.838 & 0.918 & 0.965 & 0.741 & 14.33 & 0.717 & 0.970 & 0.841 & 0.624 \\
C3          & 12.64 & 0.805 & 0.912 & 0.969 & 0.688 & 12.62 & 0.769 & 0.976 & 0.893 & 0.676 \\
C4          & 11.61 & 0.852 & 0.919 & 0.976 & 0.755 & 14.39 & 0.753 & 0.974 & 0.902 & 0.646 \\
C5          & 20.12 & 0.760 & 0.885 & 0.960 & 0.628 & 11.92 & 0.770 & 0.980 & 0.874 & 0.689 \\
C6          & 15.19 & 0.809 & 0.890 & 0.949 & 0.705 & 13.50 & 0.747 & 0.975 & 0.884 & 0.647 \\
C7          & 13.06 & 0.815 & 0.927 & 0.952 & 0.712 & 11.01 & 0.781 & 0.982 & 0.904 & 0.688 \\
C8          & 13.50 & 0.865 & 0.926 & 0.992 & 0.767 & 13.48 & 0.792 & 0.981 & 0.864 & 0.731 \\
C9          & 9.56  & 0.899 & 0.944 & 0.984 & 0.827 & 9.68  & 0.770 & 0.976 & 0.827 & 0.720 \\
clean       & 22.33 & 0.794 & 0.911 & 0.978 & 0.668 & 17.56 & 0.756 & 0.979 & 0.954 & 0.626 \\
combination & 17.55 & 0.776 & 0.883 & 0.976 & 0.643 & 12.11 & 0.744 & 0.976 & 0.886 & 0.641 \\
\bottomrule
\end{tabular}%
}
\end{table*}

\subsubsection{Analysis of Distortion Localization}

Since explicit distortion localization serves as the foundation of the proposed speech quality assessment framework, we further analyze the performance of the pretrained localization model on the test subset of the proposed partially distorted speech dataset. Table~\ref{tab:per_distortion_metrics} reports both frame-level and boundary-level evaluation metrics for different distortion categories.

Overall, the localization model achieves strong performance across diverse distortion types, yielding an overall frame-level F1 score of \textbf{0.807} with an accuracy of \textbf{0.903}. Boundary localization also demonstrates promising performance, obtaining a boundary-level F1 score of \textbf{0.742}. These results indicate that the proposed synthetic dataset provides reliable supervision for learning explicit distortion localization.

Performance varies across distortion categories according to their perceptual characteristics. Distortions that introduce clear spectral or temporal discontinuities, such as pitch shifting, clipping, MNRU, and vocoder mismatch, are localized with particularly high accuracy. In contrast, degradations with relatively smooth transitions, including reverberation, codec tandeming, and Mel-spectrogram reconstruction, are more challenging because their distortion boundaries are less distinct. This observation agrees well with human auditory perception, where distortions exhibiting abrupt spectral or temporal transitions are generally easier to localize than degradations with smooth perceptual evolution.

Despite these differences, the localization model maintains consistently high precision across almost all distortion categories but comparatively lower recall. This behavior indicates that the detector adopts a conservative localization strategy, producing few false positives while occasionally missing ambiguous distortion boundaries. Such a characteristic is desirable for the proposed speech quality assessment framework because inaccurate localization in clean regions would introduce misleading guidance during quality prediction, whereas missing a small number of ambiguous distortion frames has a much smaller impact on feature modulation.

Overall, these results demonstrate that the proposed localization model learns transferable distortion-aware representations rather than overfitting to specific distortion types. The reliable frame-level localization cues provided by the detector establish the foundation upon which the proposed DAMOS framework performs distortion-aware speech quality assessment.

\begin{table}[t]
\centering
\caption{Effect of distortion localization guidance under different mask conditions.}
\label{tab:mask_analysis}
\resizebox{\columnwidth}{!}{
\begin{tabular}{llccc}
\toprule
Dataset & Mask Condition & MSE  & SRCC  & LCC  \\
\midrule

\multirow{3}{*}{BVCC}
& Normal mask 
& 0.191 & 0.885 & 0.884 \\
& Zero mask 
& 0.191 & 0.884 & 0.884 \\
& One mask 
& 0.227 & 0.870 & 0.872 \\

\midrule

\multirow{3}{*}{TCD-VoIP}
& Normal mask 
& 0.300 & 0.837 & 0.858 \\
& Zero mask 
& 0.304 & 0.829 & 0.858 \\
& One mask 
& 0.374 & 0.818 & 0.845 \\

\bottomrule
\end{tabular}}
\end{table}

\begin{table}[!t]
  \centering
\caption{Sensitivity analysis of the modulation strength $m$ in DistortionFiLM.}
  \label{tab:gate_m}
  \begin{adjustbox}{width=0.48\textwidth}
  \begin{tabular}{ccccccc}
    \toprule
    & \multicolumn{3}{c}{\textbf{Utterance level}} & \multicolumn{3}{c}{\textbf{System level}} \\
    \cmidrule(lr){2-4} \cmidrule(lr){5-7}
    \textbf{$m$} & \textbf{MSE} & \textbf{SRCC} & \textbf{LCC} & \textbf{MSE} & \textbf{SRCC} & \textbf{LCC} \\
    \midrule
    \textbf{0.0} & \textbf{0.194} & \textbf{0.885} & \textbf{0.885} & \textbf{0.100} & \textbf{0.931} & \textbf{0.931} \\
    0.2 & 0.194 & 0.883 & 0.882 & 0.103 & 0.929 & 0.930 \\
    0.4 & 0.199 & 0.881 & 0.881 & 0.112 & 0.929 & 0.930 \\
    0.6 & 0.202 & 0.881 & 0.880 & 0.111 & 0.929 & 0.930 \\
    0.8 & 0.203 & 0.880 & 0.879 & 0.113 & 0.928 & 0.930 \\
    1.0 & 0.204 & 0.880 & 0.879 & 0.112 & 0.929 & 0.930 \\
    10.0 & 0.231 & 0.862 & 0.859 & 0.119 & 0.921 & 0.923 \\
    \bottomrule
  \end{tabular}
  \end{adjustbox}
\end{table}
\subsubsection{Sensitivity Analysis of the Modulation Strength $m$}

Although DistortionFiLM introduces explicit distortion localization information into feature learning, the relative contribution of localization guidance and original acoustic representations may influence the final quality prediction performance. To investigate this effect, we introduce a modulation bias factor $m$ to control the strength of distortion localization guidance.

Specifically, the feature modulation operation is formulated as follows:

\begin{equation}
\mathbf{x}_{e}
=
(m+\mathbf{g})
\odot
\mathbf{x}_{s},
\label{eq:modulation_strength}
\end{equation}
where $\mathbf{g}$ denotes the distortion-aware modulation gate generated from the predicted localization map, and $m$ controls the contribution of non-localized feature information. When $m=0$, the model relies entirely on the distortion localization gate for feature modulation. Increasing $m$ gradually introduces more uniformly preserved feature components, reducing the relative influence of localization guidance.

The influence of different $m$ values is evaluated on the BVCC test set, and the results are summarized in Table~\ref{tab:gate_m}. The performance remains relatively stable when $m$ is within a moderate range ($0\leq m\leq1$), demonstrating that the proposed framework is robust to variations in the balance between distortion localization and global representation preservation.

However, excessively large values of $m$ lead to noticeable performance degradation. For example, when $m$ increases to 10.0, the utterance-level SRCC decreases from 0.885 to 0.862. This degradation indicates that overly increasing the contribution of uniformly preserved features weakens the relative effect of distortion localization guidance, making the model less sensitive to perceptually important degraded regions.

Interestingly, the best performance is obtained when $m=0$, suggesting that directly relying on distortion-aware modulation can already provide effective guidance for quality prediction. Meanwhile, introducing a small bias ($m=0.2$) only causes negligible performance variation, indicating that distortion localization information acts as a complementary perceptual prior rather than a strict constraint. These results demonstrate that the proposed framework can effectively balance local distortion awareness and global quality representation learning.
\subsubsection{Mask Perturbation Analysis of Distortion Localization Guidance}

To further investigate the role of explicit distortion localization information in the proposed framework, we conduct a mask perturbation analysis on the BVCC and TCD-VoIP test sets. Specifically, the predicted distortion mask is replaced with two extreme cases: an all-zero mask, where all frames are regarded as distortion-free, and an all-one mask, where all frames are incorrectly treated as distorted. These two settings are designed to examine whether the effectiveness of the proposed distortion-aware feature modulation originates from meaningful distortion localization guidance rather than simply introducing an additional modulation branch. The results are summarized in Table~\ref{tab:mask_analysis}.

It is important to distinguish this perturbation analysis from the ablation study in Table~\ref{tab:ablation}. The latter removes DistortionFiLM entirely \emph{during training}, which directly degrades the model's optimization objective and leads to a clear performance drop (utterance-level SRCC from 0.885 to 0.877). In contrast, the mask perturbation experiment here preserves the trained DistortionFiLM module and only corrupts its \emph{input at inference time}. These two settings therefore probe different aspects of the framework: the former evaluates the necessity of localization guidance during representation learning, while the latter evaluates the sensitivity of the already-trained model to the quality of localization cues at test time. This distinction motivates the following analysis.

As shown in Table~\ref{tab:mask_analysis}, replacing the predicted mask with an all-zero mask causes only marginal degradation on both datasets (BVCC: SRCC 0.885 $\rightarrow$ 0.884; TCD-VoIP: SRCC 0.837 $\rightarrow$ 0.829), whereas the all-one mask leads to a substantially larger drop (BVCC: SRCC 0.885 $\rightarrow$ 0.870, MSE 0.191 $\rightarrow$ 0.227; TCD-VoIP: SRCC 0.837 $\rightarrow$ 0.818, MSE 0.300 $\rightarrow$ 0.374). This asymmetry can be explained directly from the formulation of the modulation gate in Eq.~\eqref{eq:film}. When $\mathbf{M}=\mathbf{0}$, the term $\mathrm{ReLU}(\mathbf{W}_1\mathbf{M})$ vanishes, and the resulting gate $\mathbf{g}=\sigma(\mathbf{0})$ collapses to a fixed, input-independent constant determined solely by the bias of $\mathbf{W}_2$. Since clean frames constitute the majority of speech content even in distorted utterances, this constant gate closely approximates the effective operating point that DistortionFiLM already adopts for most frames during training, so the resulting perturbation is mild. In contrast, setting $\mathbf{M}=\mathbf{1}$ forces the gate into a configuration that is rarely, if ever, encountered during training, since ground-truth masks in which all frames are simultaneously distorted essentially do not occur in the training distribution. This out-of-distribution input causes the modulation module to indiscriminately amplify distortion-related responses across clean and distorted regions alike, actively injecting misleading guidance rather than simply withholding it. The performance gap between the two conditions therefore reflects a difference in kind---missing localization information versus incorrect localization information---rather than a difference in degree.

This behavior also helps explain why the all-zero condition results in such limited degradation despite removing explicit localization cues altogether. Because DistortionFiLM participates in the optimization process during training, the gradient signal it provides encourages the underlying SSL representations produced by DSLA to already encode distortion-sensitive characteristics, rather than relying exclusively on the modulation gate to inject this information at test time. Consequently, part of the benefit of explicit distortion localization is absorbed into the learned representations themselves, allowing the model to retain reasonable predictive capability even when localization guidance is unavailable at inference. This is consistent with the role of DistortionFiLM as a training-time inductive bias rather than a strictly necessary runtime input, and echoes a broader pattern in learning under privileged information, where auxiliary supervision available during training continues to benefit a model even when that information is absent at test time.

Notably, this robustness profile aligns with the empirical error characteristics of the pretrained localization model reported in Table~\ref{tab:per_distortion_metrics}, where the detector consistently exhibits high precision but comparatively lower recall across distortion categories, indicating a conservative localization behavior that rarely produces false positives but occasionally misses ambiguous distortion boundaries. In practice, such recall-dominated errors resemble a partial, localized version of the zero-mask condition---i.e., sporadic omission of true distortion frames---rather than the systematic over-triggering simulated by the all-one mask. The observed robustness to the zero-mask condition is therefore not merely a favorable artifact of the perturbation experiment, but is directly relevant to real-world deployment: the proposed framework remains stable under the specific type of imperfection that the localization model is actually prone to, while the more damaging failure mode of indiscriminate over-detection is comparatively unlikely to occur given the detector's conservative precision-oriented behavior.

Taken together, these results indicate that the contribution of explicit distortion localization is twofold: it provides direct, frame-level guidance for feature modulation during inference, and it also shapes the underlying quality representations through training-time supervision, which persists even when localization information is degraded or unavailable at test time. The larger degradation observed on TCD-VoIP relative to BVCC further suggests that this benefit is more pronounced for speech signals with temporally localized degradations, such as packet loss and transmission artifacts common in communication speech, where accurate distortion localization is more consequential for quality prediction. We note that the present analysis relies on synthetic mask perturbations rather than the localization model's actual error distribution; evaluating the framework under naturally occurring localization errors remains an interesting direction for future work.
\section{Conclusion}
\label{sec:conclusion}

In this paper, we investigate the potential of explicit distortion localization as auxiliary knowledge for speech quality assessment. Different from existing approaches that rely solely on utterance-level MOS supervision, we argue that providing models with information about \textit{where distortions occur} can help them learn quality representations that better align with human perceptual mechanisms. To this end, we constructed a partially distorted speech dataset with automatically generated frame-level annotations and trained a dedicated localization model to provide transferable distortion-aware cues.

Based on the obtained localization knowledge, we proposed DAMOS, a distortion-aware speech quality assessment framework that progressively exploits explicit distortion information throughout the MOS prediction pipeline. Specifically, DSLA adaptively identifies distortion-sensitive representations, DistortionFiLM injects frame-level localization cues through distortion-guided modulation, and LQR preserves localized quality information until the final estimation. 
Extensive experiments on multiple benchmarks demonstrate that DAMOS consistently outperforms existing state-of-the-art methods and achieves strong cross-dataset generalization, confirming that explicit distortion localization provides complementary information beyond conventional MOS supervision.

Despite these promising results, several directions remain open. First, the current framework relies on synthetically generated distortion masks for both training the localization model and conducting our inference-time robustness analysis; evaluating DAMOS under the localization model's naturally occurring error distribution on real, non-synthetic distortions would provide a more direct measure of its practical robustness. Second, the localization and quality prediction models are trained separately, with the localization backbone kept frozen during MOS training; jointly optimizing distortion localization and perceptual quality estimation may allow the two objectives to reinforce each other further. Finally, extending the framework toward fully frame-level, interpretable quality assessment---where predicted distortion locations are directly exposed as human-interpretable evidence for a quality score---represents a natural next step toward more transparent and diagnostic speech quality assessment.

\bibliographystyle{IEEEtran}
\bibliography{ref}

\end{document}